\documentclass[%
 reprint,
 amsmath,amssymb,
 aps,
longbibliography]{revtex4-1}

\usepackage{graphicx}
\usepackage{dcolumn}
\usepackage{bm}
\usepackage{hyperref}
\usepackage{subfigure}
\usepackage{amsmath}
\usepackage{color}
\usepackage{pbox}
\usepackage[english]{babel}
\usepackage{siunitx}
\usepackage{gensymb}

\begin{document}


\title{$\mu$SR and Magnetometry Study of the Type-I Superconductor BeAu}

\author{J. Beare$^{1}$, M. Nugent$^{1}$, M.N. Wilson$^{1}$, Y. Cai$^{1}$, T.J.S. Munsie$^{1}$, A. Amon$^{2}$, A. Leithe-Jasper$^{2}$  Z. Gong$^{3}$, S.L. Guo$^{4}$, Z. Guguchia$^{3}$, Y. Grin$^{2}$, Y.J. Uemura$^{3}$, E. Svanidze$^{2}$ and G.M. Luke$^{1,4,5}$.}
\affiliation{ $^{1}$Department of Physics and Astronomy, McMaster University, Hamilton, Ontario, Canada L8S 4M1 \\ $^{2}$ Max-Planck-Institut f{\"u}r Chemische Physik fester Stoffe, N{\"o}thnitzer Stra{\ss}e 40, 01187 Dresden, Germany \\ $^{3}$Department of Physics, Columbia University, New York, New York 10027, USA \\ $^{4}$Department of Physics, Zhejiang University, Hangzhou 310027, China \\ $^{5}$Canadian Institute for Advanced Research, Toronto, ON, Canada M5G 1Z7 \\ $^{6}$ TRIUMF, Vancouver, BC, Canada V6T 2A3
}

\date{\today}

\begin{abstract}
We present muon spin rotation and relaxation ($\mu$SR) measurements as well as demagnetising field corrected magnetisation measurements on polycrystalline samples of the noncentrosymmetric superconductor BeAu. From $\mu$SR measurements in a transverse field, we determine that BeAu is a type-I superconductor with H$_{c} =$ 256 Oe, amending the previous understanding of the compound as a type-II superconductor. To account for demagnetising effects in magnetisation measurements, we produce an ellipsoidal sample, for which a demagnetisation factor can be calculated. After correcting for demagnetising effects, our magnetisation results are in agreement with our $\mu$SR measurements. Using both types of measurements we construct a phase diagram from T = 30 mK to T$_{c} \approx$ 3.25 K. We then study the effect of hydrostatic pressure and find that 450 MPa decreases T$_{c}$ by 34 mK, comparable to the change seen in type-I elemental superconductors Sn, In and Ta, suggesting BeAu is far from a quantum critical point accessible by the application of pressure.
\end{abstract}

\maketitle

\section{\label{sec:level1}Introduction}

The absence of inversion symmetry in noncentrosymmetric superconductors results in an antisymmetric spin-orbit coupling, which splits otherwise degenerate electronic bands \cite{Bauer2012}. Parity of the superconducting parameter is no longer conserved, allowing spin-singlet and spin-triplet states to mix. These mixed parity states are generally expected to give rise to point or line nodes in the superconducting gap \cite{Hayashi2006,Sigrist2007,Takimoto2009}.

Evidence for line nodes has been found in CePt$_{3}$Si \cite{Bonalde2005}, CeIrSi$_{3}$ \cite{Mukuda2008}, Mg$_{10}$Ir$_{19}$B$_{16}$ \cite{Bonalde2009}, Mo$_{3}$Al$_{2}$C \cite{Bauer2010}, and Li$_{2}$Pt$_{3}$B \cite{Nishiyama2007},  while many other noncentrosymmetric superconductors possess fully gapped states \cite{Yuan2006,Isobe2016,Anand2011,Barker2015,Hillier2009,Iwamoto1998}. Other noncentrosymmetric superconductors such as La$_{2}$C$_{3}$ \cite{Sugawara2007} and TaRh$_{2}$B$_{2}$ exhibit multigap behaviour. A detailed analysis of the possible pairing mechanisms \cite{Samokhin2008} shows that either isotropic or nodal gaps are possible, depending on the anisotropy of the pairing mechanism. The anisotropy may depend upon the strength of spin-orbit coupling in the material \cite{Yuan2006}. 

In addition to breaking parity symmetry, noncentrosymmetric superconductors can also exhibit time-reversal symmetry breaking. Muon spin rotation and relaxation ($\mu$SR) has been used to detect time reversal symmetry breaking in the noncentrosymmetric superconductors Re$_{6}$D (D = Zr, Hf, Ti) \cite{Khan2016,Singh2016,Singh2018}, La$_{7}$Ir$_{3}$ \cite{Barker2015}, LaNiC$_{2}$ \cite{Hillier2009} and SrPtAs \cite{Biswas2013} as well as several centrosymmetric superconductors \cite{Luke1998,Xia2006,Luke1993,Dereotier1995, Aoki2003,Aoki2007,Yarzhemsky2009,Bhattacharyya2015}. Further study of noncentrosymmetric superconductors, such as BeAu, is required to study the unconventional pairing mechanisms and the diverse gap properties that have been exhibited.

BeAu exhibits cubic space group symmetry of \emph{P2$_{1}$3} (FeSi) with lattice parameter a = 4.6699(4) \si{\angstrom}. It exhibits conventional Bardeen-Cooper-Schrieffer (BCS) type superconductivity below T$_{c} \approx$ 3.3 K with a full Meissner flux expulsion, resisitivity drop and the specific heat jump at the superconducting transition is equal to the normal state specific heat, indicating bulk superconductivity. BeAu can be classified as a weakly coupled superconductor with $\Delta C / \gamma_{n} $T$_{c}\approx$ 1.26, $\lambda_{e-p}$ = 0.5, and $2\Delta \left( 0 \right) / k_{B}$T$_{c}$ = 3.72. Previously BeAu was thought to be a type-II superconductor with lower critical field H$_{c1} =$ 32 Oe and upper-critical field H$_{c2} =$ 335 Oe based upon magnetisation measurements which used a spherical demagnetisation factor N = $\frac{1}{3}$ when measuring an irregularly shaped sample. This is a common first assumption when correcting for demagnetising effects that we would like to improve in this work. Zero-field $\mu SR$ measurements presented in Ref. \cite{Amon2018} show no time reversal symmetry breaking.

$\mu$SR is a powerful technique which can measure internal fields due to time-reversal symmetry breaking, give an accurate measurement of the penetration depth and coherence length of type-II superconductor. As muons are a local probe of magnetism, $\mu$ SR is insensitive to the large demagnetising fields produced outside of superconductors. In this work, transverse field $\mu$SR measurements are used to demonstrate that BeAu is a type-I superconductor with H$_{c} =$ 256 Oe. Demagnetising effects in superconductors can cause parts of a sample to experience a magnetic field that is larger than the applied field. If the local field due to demagnetising effects at the surface of a type-I superconductor is above H$_{c}$, then at least part of the sample enters the normal state. In this situation, the free energy is minimized by a complicated structure of alternating normal and superconducting regions which depends upon the geometry of the sample, as well as the coherence length and penetration depth \cite{deGennes1999,Tinkham1975, Landau1948}. This is known as the intermediate state of a type-I superconductor and occurs for applied fields H$_{a}$ between (1 - N) H$_{c}$ and H$_{c}$, where N is the demagnetisation factor of the sample. At low applied fields, close to (1 - N) H$_{c}$, the free energy of the intermediate state is minimized when the normal regions of the sample have an internal field of H$_{c}$. At high magnetic fields, close to H$_{c}$, interface effects between normal and superconducting regions, as well as surface effects, modify the thermodynamic critical field to the slightly reduced value H$_{cI}$ given by Eq. \ref{eq:IntermediateField} \cite{deGennes1999,Tinkham1975, Landau1948}. The magnetic moment of a muon landing within a type-I superconductor will therefore either be stationary in the superconducting regions (where the field is zero) or will precess in a field very close to H$_{c}$ in the normal regions. The probability a muon experiences either field is equal to the volume fraction of the sample in the respective states and therefore a $\mu$SR measurement directly measures the superconducting volume fraction of the sample. $\mu$SR has been used to study elemental type-I superconductors such as Sn (IV) \citep{Egorov2001} and we find qualitatively similar results. The most striking indication of type-I behaviour in BeAu is that the normal regions of the sample have a relatively constant value of the internal field of H$_{c} = $ 256 Oe, for all applied fields between 50 Oe and 250 Oe. This can be contrasted with the expected internal field behaviour of the vortex state of a type-II superconductor where the internal field in the normal cores is always less than the applied field and, in general, changes as a function of the applied field. BeAu joins a relatively small list of non-elemental compounds which show type-I behaviour \cite{Svanidze2012, Yamada2018, Bekaert2016, Salis2018, Gottlieb1992, Kriener2008, Anand2011, Smidman2014, Wakui2009}.
Our $\mu$SR results call for a more accurate accounting of demagnetising effects in magnetisation measurements in order to reconcile previous magnetisation measurements \citep{Amon2018} with the identification of BeAu as a type-I superconductor. To this end, a sample of BeAu is shaped into an ellipsoid so that demagnetising effects can be accounted for more accurately than is typically done in magnetisation studies of superconductors. Using this geometry a demagnetising factor for the sample can be calculated which allows us to determine the internal field of the normal regions and  demonstrate the type-I nature of this compound with magnetisation measurements. 

Pressure can often have a dramatic effect on superconductivity, especially when the system is near a quantum critical point. Cerium based noncentrosymmetric superconductors have shown unusual behaviour under the application of pressure with a pressure induced superconducting transition discovered in CeIrGe$_{3}$, CeCoGe$_{3}$ and CeRhSi$_{3}$ \cite{Honda2010,Settai2007,Kimara2005}, while  CePt$_{3}$Si shows a strongly decreasing critical temperature as a function of applied pressure with a complete loss of superconductivity at 1.5 GPa \cite{Yasuda2004}. Iron pnictide superconductors have also shown exotic behaviour under the application of pressure \cite{Mizuguchi2008}. This can be compared to elemental type-I superconductors which have a small decrease in T$_{c}$ with the application of pressure but no exotic behaviour is observed \cite{Jennings1959}. To compare BeAu to these cases we study the effect of hydrostatic pressure and find that at 450 MPa BeAu still exhibits type-I behaviour with a decrease in T$_{c}$ of 44 mK. This decrease in T$_{c}$ is similar to the change seen in type-I elemental superconductors such as Sn, In and Ta under the same conditions \cite{Jennings1959} and suggests that BeAu is far from a quantum critical point that can be accessed by applying pressure.

\section{\label{sec:level1}Experimental Methods}

Polycrystalline samples were synthesized by arc melting from elemental Be (Heraeus, $\geq$99.9 wt.$\%$) and Au (Alfa Aesar, $\geq$99.95 wt.$\%$) in a 51:49 ratio, with mass loss of less than 0.3$\%$. A small excess of beryllium was added in order to compensate for the Be loss due to evaporation. After melting, the boule was annealed in an argon atmosphere for 48 hours at 400  $\degree$C. Further details can be found in Ref. \cite{Amon2018}. 

Transverse field (TF) Muon Spin Rotation and Relaxation ($\mu$SR) measurements were performed on the M15 and M20 beamlines at the TRIUMF Laboratory in Vancouver, Canada. A spectrometer incorporating a dilution refrigerator was used on the M15 beamline which allows for measurements in the temperature range of 0.025-10 K. The experimental set-up makes use of a superconducting magnet to allow for fields up to 5 T. Samples were mounted on a silver cold finger to ensure good thermal conductivity and to give a well defined $\mu$SR background signal with minimal relaxation. The instrument has a time resolution of 0.4 ns. The field was applied parallel to the direction of the muon beam and measurements were taken with the initial muon spin direction perpendicular to the field (TF). Several approximately elliptical discs of BeAu, each 0.5 mm thick and roughly 2.75 mm x 3.75 mm, were mounted on the cold finger such that a large fraction of the muon beam spot was covered. The direction of the applied field was perpendicular to the flat faces of the samples. Copper coil electromagnets were used to compensate for any stray fields. The LAMPF spectrometer was used on the M20 beamline which allows measurements in the temperature range from 2-300 K in an applied field up to 0.4 T. A silver cold finger is not required on LAMPF and the background signal is greatly reduced. Thin aluminium backed mylar was used to mount a mosaic of BeAu discs were mounted to a copper square cut-out. The experiment on M20 was performed using a similar transverse field setup as on M15. The $\mu$SRfit software package was used to analyse the $\mu$SR data \cite{Suter2012}.

Magnetometry measurements were taken at McMaster University using a Quantum Design XL-5 MPMS which allows for measurements down to 1.8 K. We used a GC10/3 gas pressure cell from the Institute of High Pressure Physics, Polish Academy of Sciences, inserted into the MPMS to allow for magnetometry measurements under pressure up to 700 MPa, above 2K. A 102.5 mg, nearly ellipsoidal sample with a = b = 2.75 mm and c = 1.90 mm was produced through grinding with a spherically concave, diamond Dremel head in a fume hood while submerged in mineral oil. A small section on either side of the c-axis is flat due to the constraint of grinding while submerged in mineral oil to avoid the dispersal of toxic Be/BeO dust. The sample was measured in the MPMS at ambient pressure after which it needed to be ground down (under the same conditions) to a = b = 2.69 mm and c = 1.86 mm to fit inside the 3 mm diameter pressure cell.

\section{\label{sec:level1}Results and Discussion}

\begin{figure}[!h]
\centering
\subfigure[\label{fig:75G_Asym}]{
\includegraphics[width=0.4\textwidth]{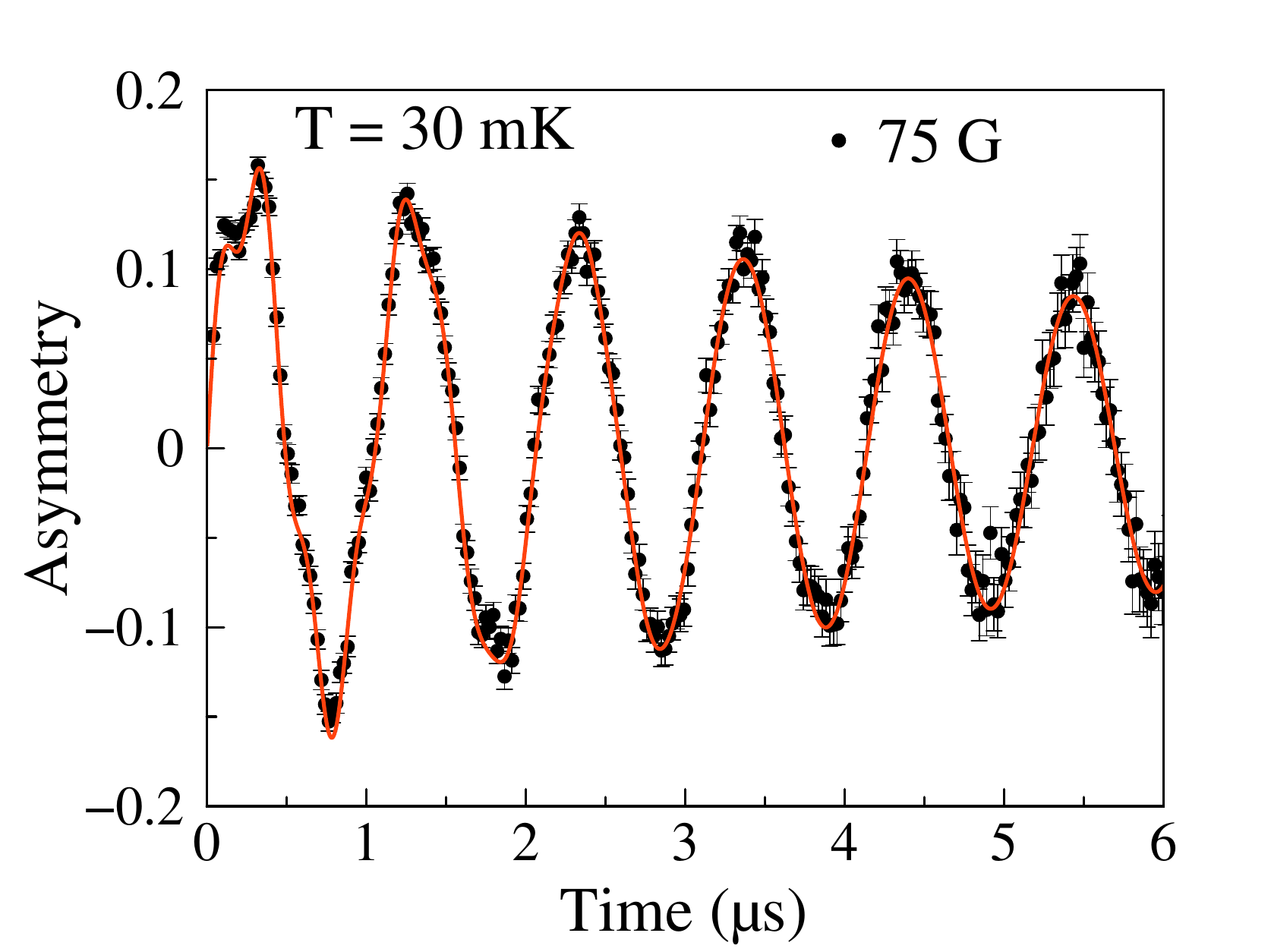}}
\subfigure[\label{fig:75G_M20}]{
\includegraphics[width=0.4\textwidth]{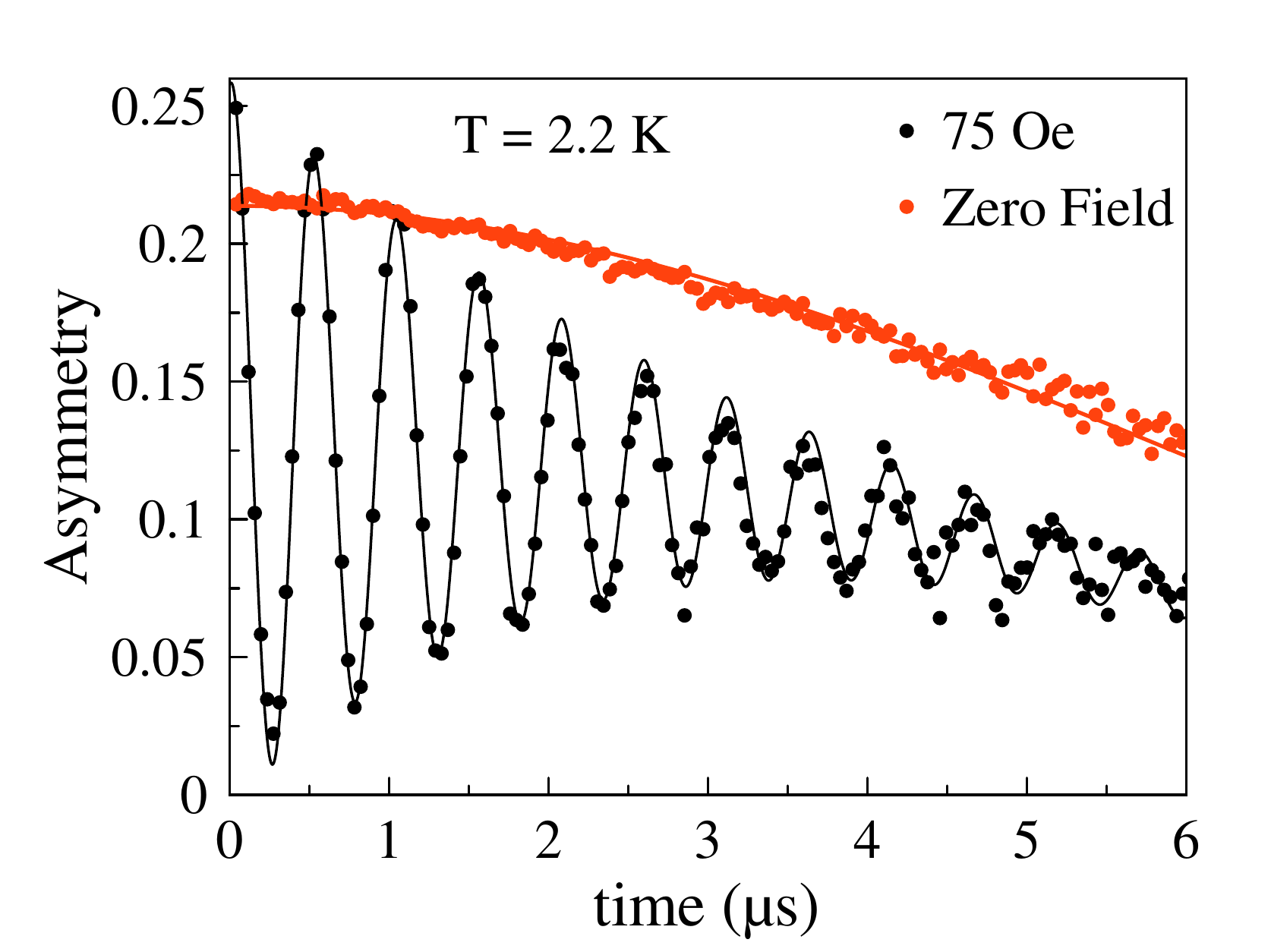}}
\caption{\label{fig:75G_2p2Asym} (a) BeAu $\mu$SR Asymmetry at 30 mK on the M15 beamline in an applied transverse field of 75 G. (b) $\mu$SR Asymmetry at 2.2 K on the M20 beamline in an applied transverse field of 75 G (Black) and zero field (red). The non-oscillating component of the 75 G data relaxes with the same rate as the zero field data, showing that there are regions in the sample at zero field}
\end{figure}

Fig. \ref{fig:75G_Asym} shows the transverse field (TF) $\mu$SR asymmetry spectrum at 30mK on the M15 beamline  while \ref{fig:75G_M20} shows the asymmetry spectrum on M20 at 2.2 K (black circles), both after field cooling in an applied field of 75 G. The M15 data show large oscillations as expected for muons precessing in an applied transverse field with additional contributions coming from a fraction of muons precessing in a different field. This is due to a significant fraction of muons stopping in the silver cold finger and precessing in the applied magnetic field while another fraction of muons stop in the sample and precess in the local field of the sample. The M15 data show a reduced initial asymmetry, A$_{0}$, while in a superconducting state due to there being no detectors in the direction of the initial muon polarization when using the TF geometry on M15. Muons landing in the superconducting regions of the sample will not precess and their decay products, which tend to travel in the direction of the muon spin polarization, will be less likely to be detected. This generates a "missing asymmetry". In the intermediate state of a type-I superconductor the $\mu$SR asymmetry measures the total fraction of the signal coming from the sample, $\emph{F}$, as well as from superconducting regions in the sample, F$_{S}$, and the asymmetry spectrum on the M15 beamline may be fit to 
\begin{multline}\label{eq:AsymmetryM15}
A(t) = A_{0} \lbrace F\left(1-F_{S}\right)\cos\left(\gamma_{\mu} H_{N} +  \phi\right) \exp\left(-\frac{1}{2}\left( \sigma_{N} t \right)^{2} \right) \\ + \left(1-F \right) \cos \left( \gamma_{\mu} H_{bkg} + \phi \right)\exp\left( -\lambda_{bkg} t \right)  \rbrace ,
\end{multline} 
where $A_{0}$ is the initial asymmetry, H$_{N}$ the internal field in the normal regions of the sample, H$_{bkg}$ the background field seen by muons stopping outside of the sample, $\phi$ the phase shift in the $\mu$SR asymmetry, $\sigma_{N}$ the relaxation rate of the normal regions of the sample and $\sigma_{bkg}$ the relaxation rate muons stopping outside of the sample. As $\mu$SR randomly samples the entire volume of the material, the fraction of the signal coming from normal/superconducting regions is equivalent to the volume fraction of the sample in the normal/superconducting state. To determine A$_{0}$ and to account for extrinsic effects, high temperature (above T$_{c}$) data was taken for each applied field and fits to Eq. \ref{eq:AsymmetryM15} were performed after fixing F$_{S}$ = 0. The missing asymmetry when comparing low temperature and high temperature measurements allows F$_{S}$ to be determined on the M15 beamline. A fit to Eq. \ref{eq:AsymmetryM15} at 30 mK and 75 G is given by the solid line in Fig. \ref{fig:75G_Asym}.

LAMPF on M20 has detectors in the direction of the initial muon polarization and the 75 Oe data (Fig. \ref{fig:75G_M20}) shows a large fraction of the signal is very slowly relaxing, which is expected if a large fraction of the signal is coming from superconducting regions with zero field. The relaxation rate of this non-oscillating component is equal to the relaxation rate of measurements performed in zero field and so this component may be identified with regions of the sample that are superconducting. The visual difference in relaxation rate between the 75 G and zero field measurements in Fig. \ref{fig:75G_M20} is due to the different initial asymmetries of the two types of measurements. The oscillating component of the signal comes from normal regions of the superconductor that are experiencing a non-zero field. There is very little background signal on M20 and the M20 asymmetry may be fit to
\begin{multline}\label{eq:AsymmetryM20}
A(t) = A_{0} \lbrace \left(1- F_{S}\right)\cos\left(\gamma_{\mu} H_{N} +  \phi\right) \exp\left(-\lambda_{N} t  \right) + \\ F_{s} \exp\left(-\frac{1}{2}\left( \sigma_{N} t \right)^{2} \right)  \rbrace,
\end{multline}
where A$_{0}$ is the initial asymmetry, F$_{S}$ is the superconducting volume fraction, H$_{N}$ and $\lambda_{N}$ are the internal field and relaxation in the normal regions of the sample, $\phi$ the phase shift of the normal regions, H$_{M}$ and $\lambda_{M}$ the internal field and the relaxation rate of the superconducting regions of the sample and $\phi_{M}$ the phase shift of the superconducting regions. A fit to Eq. \ref{eq:AsymmetryM20} at 2.2 K and 75 G is given by the solid line in Fig.\ref{fig:75G_2p2Asym}. H$_{M}$ consistently fits to values close to zero while in the intermediate state. To determine the expected internal field and superconducting fraction a discussion of type-I superconductors is required.

For applied fields, H$_{a}$ above (1 - N) H$_{c}$, where N is the demagnetisation factor of the sample, demagnetising fields cause regions of the surface of a type-I superconductor to experience a field that is larger than the critical field H$_{c}$; inevitably causing parts of the sample to enter the normal state. For applied fields H$_{c}\left( 1 - N \right) \leq $ H$_{a} \leqslant $ H$_{c}$ a type-I superconductor will have a complicated structure of coexisting superconducting and normal regions known as the Intermediate State. Just above approximately (1 - N)H$_{c}$, it can be shown that equilibrium between the superconducting and normal phases can only be achieved if the normal regions have an internal magnetic field of H$_{c}$ \cite{deGennes1999,Tinkham1975, Landau1983,Andrew1948}. The field changes from 0 in the superconducting state to $H_{c}$ over a distance $\delta \approx \xi - \lambda_{L}$ where $\xi$ is the coherence length and $\lambda_{L}$ the London penetration depth, which requires energy. There is also a surface energy associated with maintaining the normal regions at H$_{c}$ in an applied field of H$_{a}$. As the applied field increases more normal regions are generated, increasing the energy requirements of both affects that slightly reduces the thermodynamic critical field from H$_{c}$ to the intermediate critical field H$_{cI}$. In the case of a thin plate oriented perpendicular to H$_{a}$, under the  assumption of a laminar domain structure of normal and superconducting regions, H$_{cI}$ is approximated by
\begin{equation}\label{eq:IntermediateField}
H_{cI} \approx H_{c} \left[ 1-2\theta \left(\frac{\delta}{d}\right)^{1/2}\right], \theta = \sqrt{\frac{\ln 2}{\pi}},
\end{equation}
where $d$ is the plate thickness, $\delta$ the thickness of the interface between superconducting and normal regions and $\theta$ is a numerical constant which depends on the assumed domain structure of the superconducting-normal regions as well as the geometry of the sample \cite{deGennes1999,Tinkham1975,Landau1948}. Similar effects will also slightly raise the Meissner to Intermediate state transition field from (1 - N)H$_{c}$ \cite{Landau1948}.

To complicate matters more, the structure of superconducting-normal domains which minimizes the free energy changes as a function of H$_{a}$. At applied fields just above $\approx$ H$_{cI} \left( 1 - N \right)$ the free energy is minimized by having tubular-threadlike normal regions pierce the superconductor, while at intermediate fields the free energy is minimized by having corrugated laminae of superconducting and normal layers. Finally at fields close to H$_{cI}$ tubular-threadlike superconducting regions pierce normal metal \cite{deGennes1999,Tinkham1975,Landau1948,Landau1983,Fortini1972}.This behaviour has been observed before \cite{Livingston1963_1,Livingston1963_2,Huebener1979} but, the exact behaviour of a material in the intermediate state is hard to predict as, it has been shown that \emph{a priori} the free energy differences of various spatial configurations of the intermediate state are quite small. Observationally, the spatial configuration selection depends upon the exact experimental conditions, as well as sample quality \cite{Huebener1979,Tinkham1975, Landau1983}. 

The internal field in the normal regions of a type-I superconductor remain relatively constant at H$_{c}$ until an applied field comparable with H$_{c}$ is reached. A transverse field $\mu$SR experiment on a type-I superconductor, in an applied field above H$_{c}\left( 1 - N \right)$ and sufficiently far below H$_{c}$, will therefore show muons in the normal regions of the sample precessing with a frequency distribution centered around $\omega = \gamma_{\mu} H_{c}$ where $\gamma_{\mu}$= 135 MHz/T is the gyromagnetic ratio of the muon \cite{Egorov2001}. This can be contrasted with the expected results of a transverse field $\mu$SR experiment on a type-II superconductor in the vortex state where muons will precess at a frequency $\omega = \gamma_{\mu} H_{int}$ where H$_{int}$ is an asymmetric distribution falling entirely below H$_{a}$ \cite{Wilson2017}. It should also be noted that while a type-I superconductor in the intermediate state will have an approximately constant internal field with H$_{int} \approx$ H$_{c}$ or H$_{int}$ = 0 (normal versus superconducting regions) for all applied fields in the range H$_{c}\left( 1 - N \right) \leq $ H$_{a} \leq $ H$_{c}$, the distribution of H$_{int}$ for a type-II superconductor in the vortex state will generally depend upon H$_{a}$.

\begin{figure}[!h]
\centering
\includegraphics[width=0.37\textwidth]{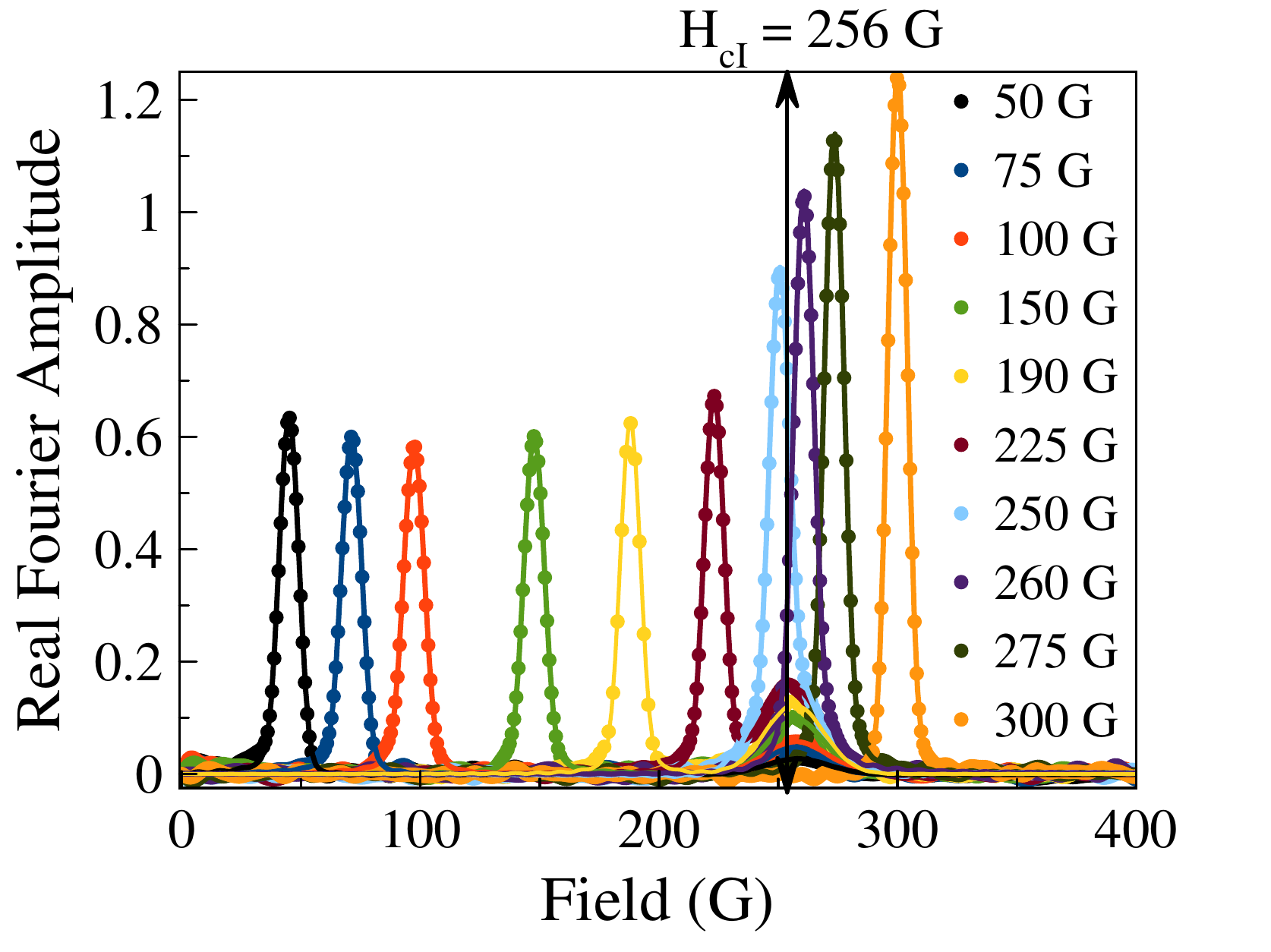}
\caption{\label{fig:muSR_Fourier} The Fourier transform of the M15 $\mu$SR Asymmetry for a variety of fields at 30 mK. The sharp peaks in the data as well as color indicate the applied field experienced by muons stopping in the silver cold finger. The broad peak centered around 256 Oe for all applied fields below H$_{a}$ = 260 Oe unambiguously shows BeAu to be a type-I superconductor}
\end{figure}

The Fourier transform  of the $\mu$SR spectra gives the probability distribution of H$_{int}$ in both the intermediate state of a type-I and the vortex state of a type-II superconductor. The probability distribution for both types of superconductors will show a sharp background peak at $H_{a}$ on the M15 beamline from muons stopping in the silver cold finger but the internal field in a type-I superconductor will not change as a function of applied field whereas as type-II superconducting will have a varying internal field. The short, broad peak centered at 256 Oe for all applied fields H$_{a} \leq$ 260 G in Fig. \ref{fig:muSR_Fourier} unambiguously demonstrates that BeAu is a type-I superconductor with H$_{c} \approx$ 256 Oe.  

\begin{figure}[!h]
\centering
\subfigure[\label{fig:MeisnnerVsB}]{
\includegraphics[width=0.37\textwidth]{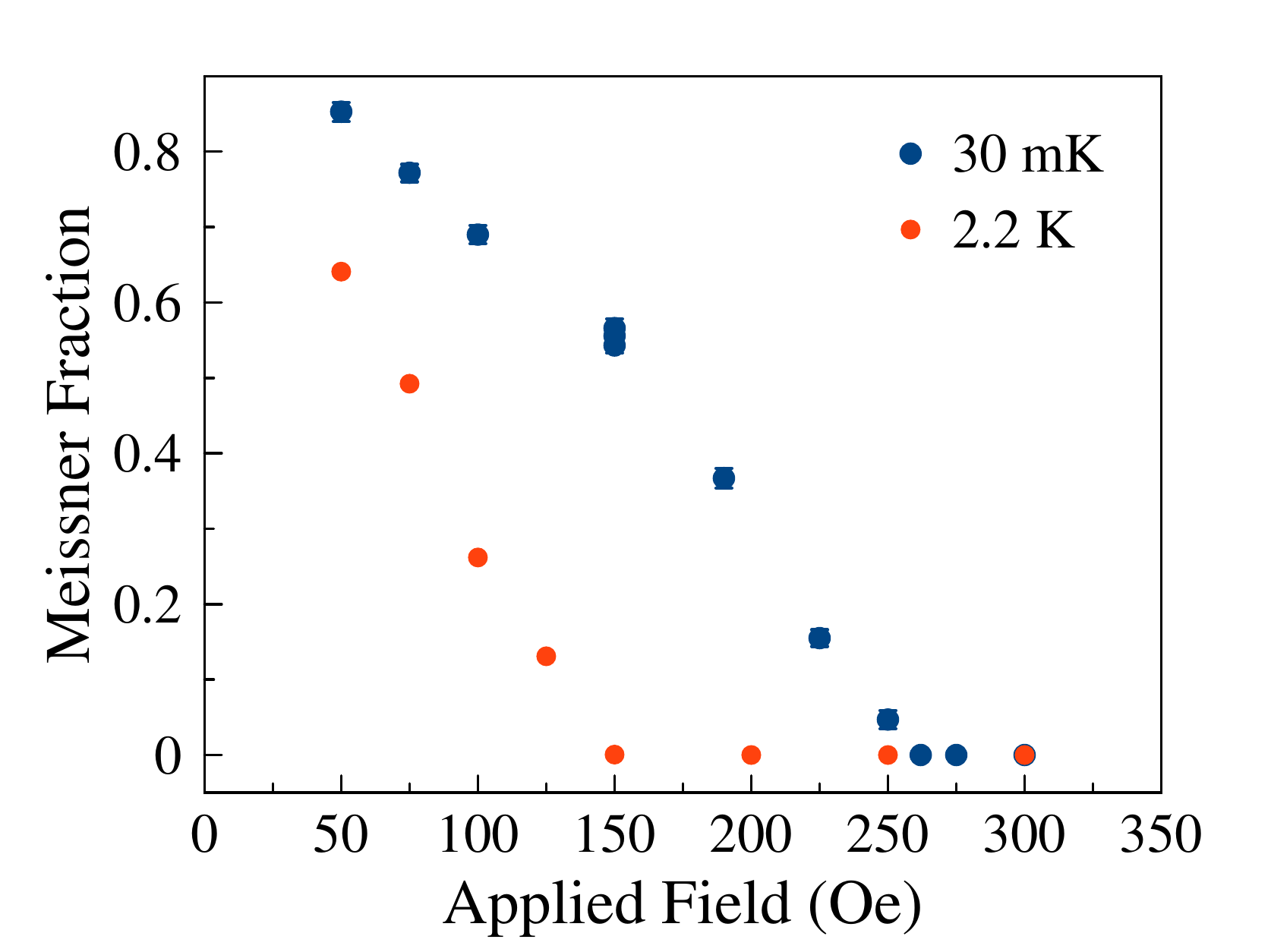}}
\subfigure[\label{fig:InternalFieldVsB}]{
\includegraphics[width=0.37\textwidth]{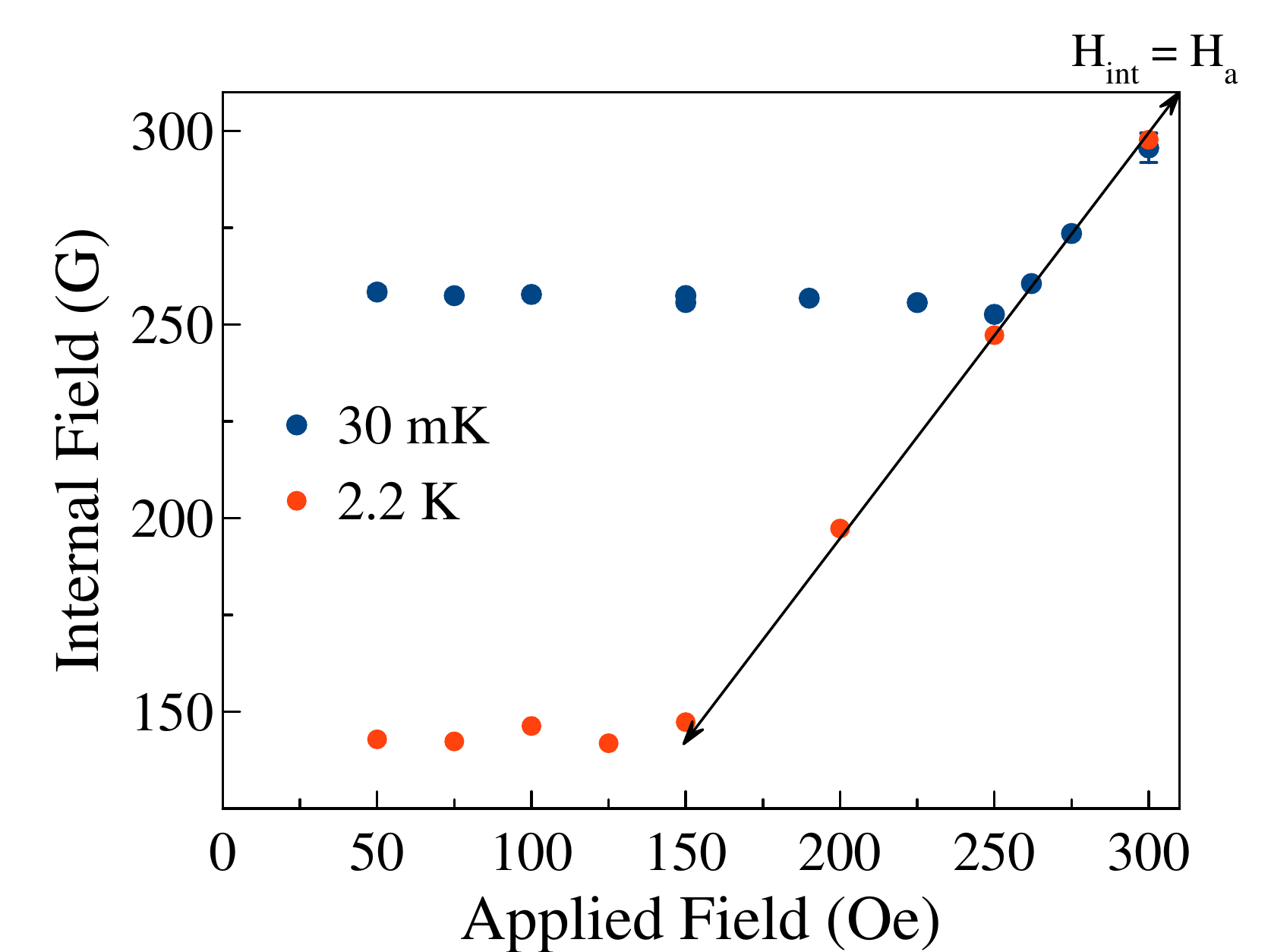}}
\caption{\label{fig:fig2} (a) The Meissner fraction as a function of applied field at 30 mK on M15 (blue) and 2.2 K on M20 (red). The Meissner fraction increases as the field decreases as is expected for a type-I superconductor in the intermediate state. Where not shown, uncertainties are smaller than marker size. (b) The internal field in the normal regions of the sample as a function of applied field at 30 mK on M15 (blue) and 2.2 K on M20 (red). The data shows the normal regions have an aproximately constant internal field while in the intermediate state. For applied fields above H$_{cI}$, H$_{Int} = $H$_{a}$ indicating the sample is in the normal state.}
\end{figure}

The results from fitting the 30 mK data at various applied magnetic fields in the time domain are shown in Fig.\ref{fig:fig2}. The volume fraction of the sample in the Meissner state of a type-I superconductor should decrease from 100$\%$ near (1 - N)H$_{c}$ approximately linearly for low fields \cite{Tinkham1975}. As the applied field increases and approaches H$_{cI}$, the superconducting volume fraction picks up a small correction given by Eq. 2.28 of Ref. \cite{Tinkham1975}. The Meissner fraction for BeAu given in Fig. \ref{fig:MeisnnerVsB} shows qualitatively similar behaviour to \cite{Egorov2001} and follows the expected linear behaviour for fields not too close to H$_{c}$ \cite{Tinkham1975,deGennes1999}. Fig.\ref{fig:InternalFieldVsB} shows that the internal field in the normal regions of the sample is nearly constant as a function of H$_{a}$ until H$_{a}\approx$ H$_{cI}$ is reached. The variation in internal field is due to the contribution of the interface energy between normal and superconducting regions as well as the interface energy between normal regions and regions outside the sample as a significant fraction of the sample enters the normal state \cite{deGennes1999,Tinkham1975,Landau1948, Landau1983, Fortini1972}.    

Our data show that BeAu is in the intermediate state from 50-256 Oe with some variability in internal field due to interface energy and a changing normal-superconductor structure from 200-256 Oe. The transition to the normal state above $\approx$ 256 Oe is indicated in Fig. \ref{fig:MeisnnerVsB} and indirectly shown in Fig. \ref{fig:InternalFieldVsB} where H$_{int}$ = H$_{a}$ above 256 Oe. An approximate demagnetisation factor of the measured discs is taken from a table \cite{Osborn1945} and gives N = 0.82. Using H$_{M-I} \approx H_{cI} \left( 1 - N) \right)$ shows the Meissner state is expected for H$_{a}$ less than $\approx$ 46 Oe. As we did not measure below 50 Oe, a pure Meissner state is not seen in our data. 

\begin{figure}[!h]
\centering
\subfigure[\label{fig:MeisnnerFractionVsT}]{
\includegraphics[width=0.37\textwidth]{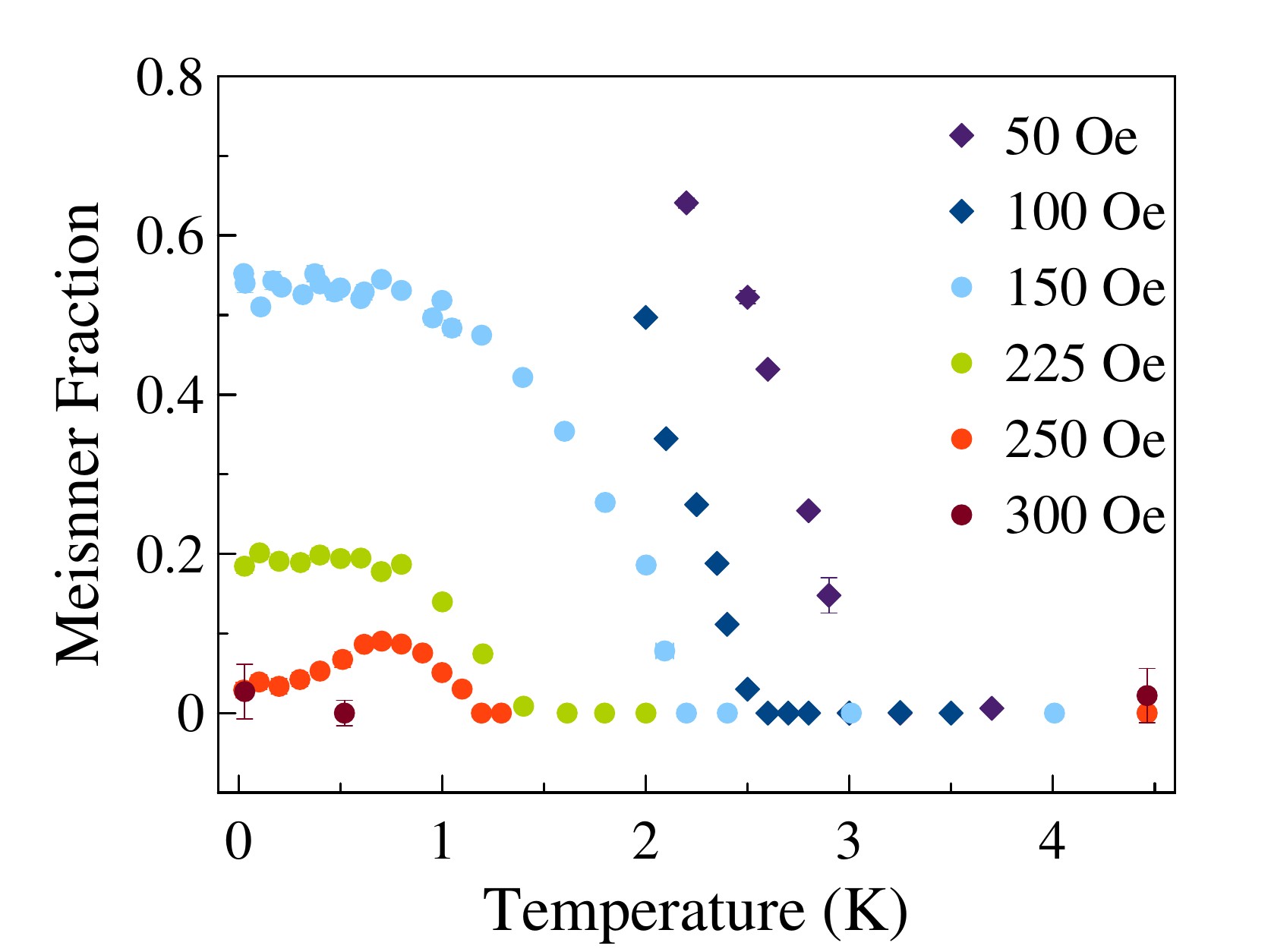}}
\subfigure[\label{fig:InternalFieldVsT}]{
\includegraphics[width=0.37\textwidth]{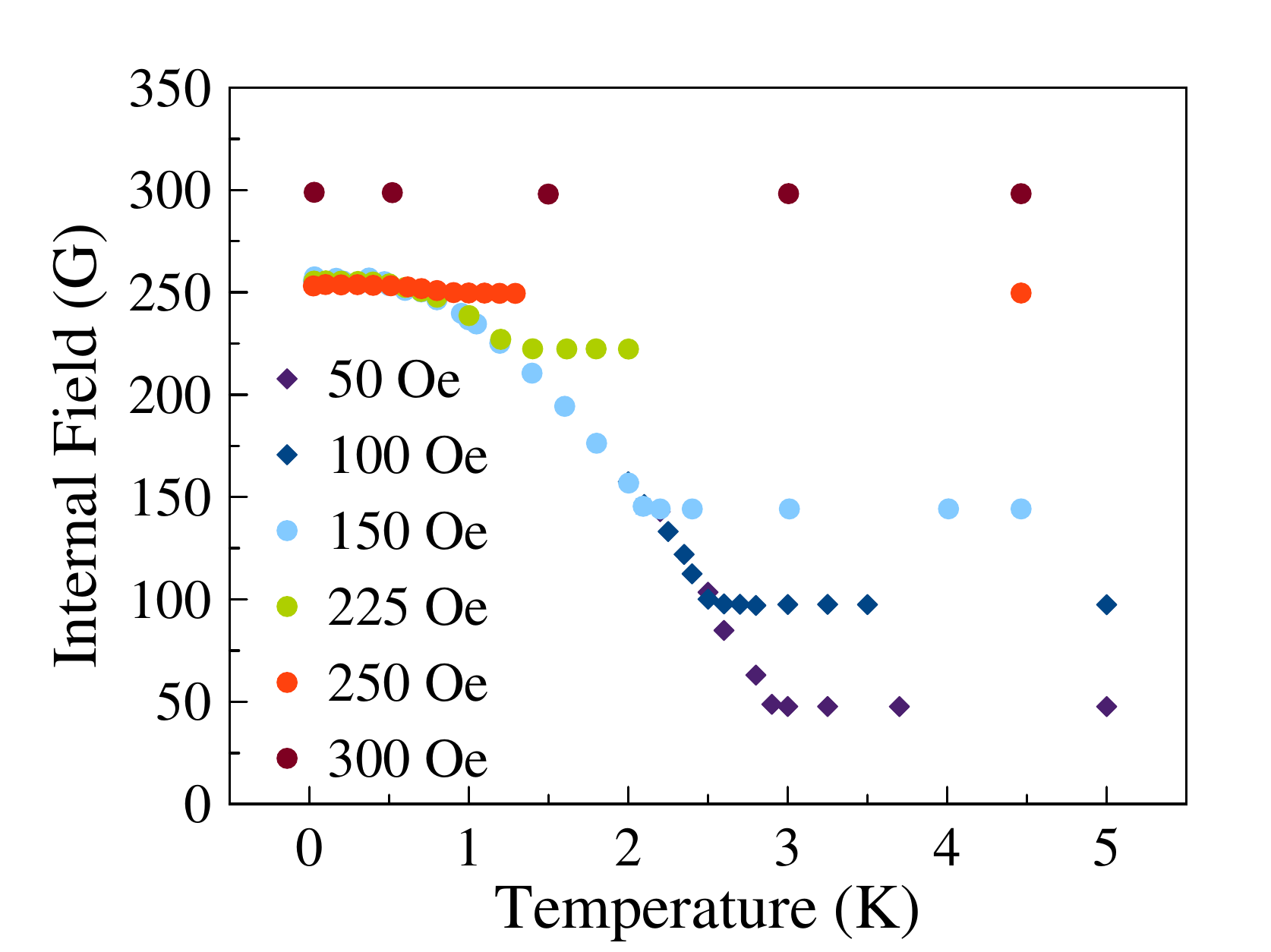}}
\caption{\label{fig:fig3} (a) Meissner fraction state as a function of temperature. The Meissner fraction increases as the temperature decreases for fields below H$_{cI}$ as is expected for a type-I superconductor in the intermediate state. Near H$_{cI}$, restructuring of the superconducting-normal regions leads to a more complicated behaviour of the Meissner fraction. Uncertainties are smaller than marker size where not shown (b) The internal field in the normal regions of the sample as a function of temperature. For an applied field below H$_{cI}$(T), the internal field is equal to approximately H$_{c}$(T) and $\mu$SR is able to measure H$_{c}$(T) while in the intermediate state. For applied fields above H$_{c}$(T) the sample is in the normal state and the internal field is equal to the applied field}
\end{figure}

The $\mu$SR asymmetry of BeAu was also studied as a function of temperature for the applied fields 50 G, 100 G, 150 G, 225 G, 250 G, 300 G and results are shown in Fig. \ref{fig:fig3}. The fraction of the sample in the Meissner state generally increases as temperature decreases (Fig. \ref{fig:MeisnnerFractionVsT}) as is expected in a type-I superconductor. The exception to this is the 250 G data set which is close enough to H$_{c}$ that the superconducting-normal domain structure is likely to change as a function of temperature, affecting the relative size of the normal and superconducting volumes. The internal field in the normal regions as a function of temperature is shown in Fig. \ref{fig:InternalFieldVsT} and demonstrates that the $\mu$SR technique can be used to trace out a partial H versus T phase diagram for type-I superconductors while in the intermediate state, i.e. when H$_{cI}$(1 - N)$\leq H_{a}\leq$H$_{cI}$.

Our $\mu$SR results unambiguously show that BeAu is a type-I superconductor, motivating a new careful study of the magnetisation properties of BeAu using a well defined sample geometry so that demagnetising effects can be accounted for more accurately than is typically done in studies of superconductors. For a type-I superconductor of an ellipsoidal shape the field at the surface of the equator of the sample, H$_{eq}$, when measuring in an applied field perpendicular to this equator, is given by
\begin{equation}\label{Eq:Demag}
H_{eq} = H_{a} - 4\pi NM,
\end{equation}
where $N$ is the demagnetising factor and $M$ the magnetisation of the sample \cite{Tinkham1975,Landau1983,deGennes1999}. An ellipsoid of revolution is one of the few shapes for which the demagnetisation factor can be calculated analytically which is why we produced a sample with this geometry\cite{Osborn1945,Beleggia2006}. The surface field at the equator of a superconducting ellipsoid is the maximum field that the sample experiences and is given by,
\begin{equation}
H_{eq} = \dfrac{H_{a}}{1 - N}.
\end{equation}\cite{Tinkham1975,deGennes1999}
\begin{figure}[!h]
\centering
\subfigure[\label{fig:MvsHRaw}]{
\includegraphics[width=0.37\textwidth]{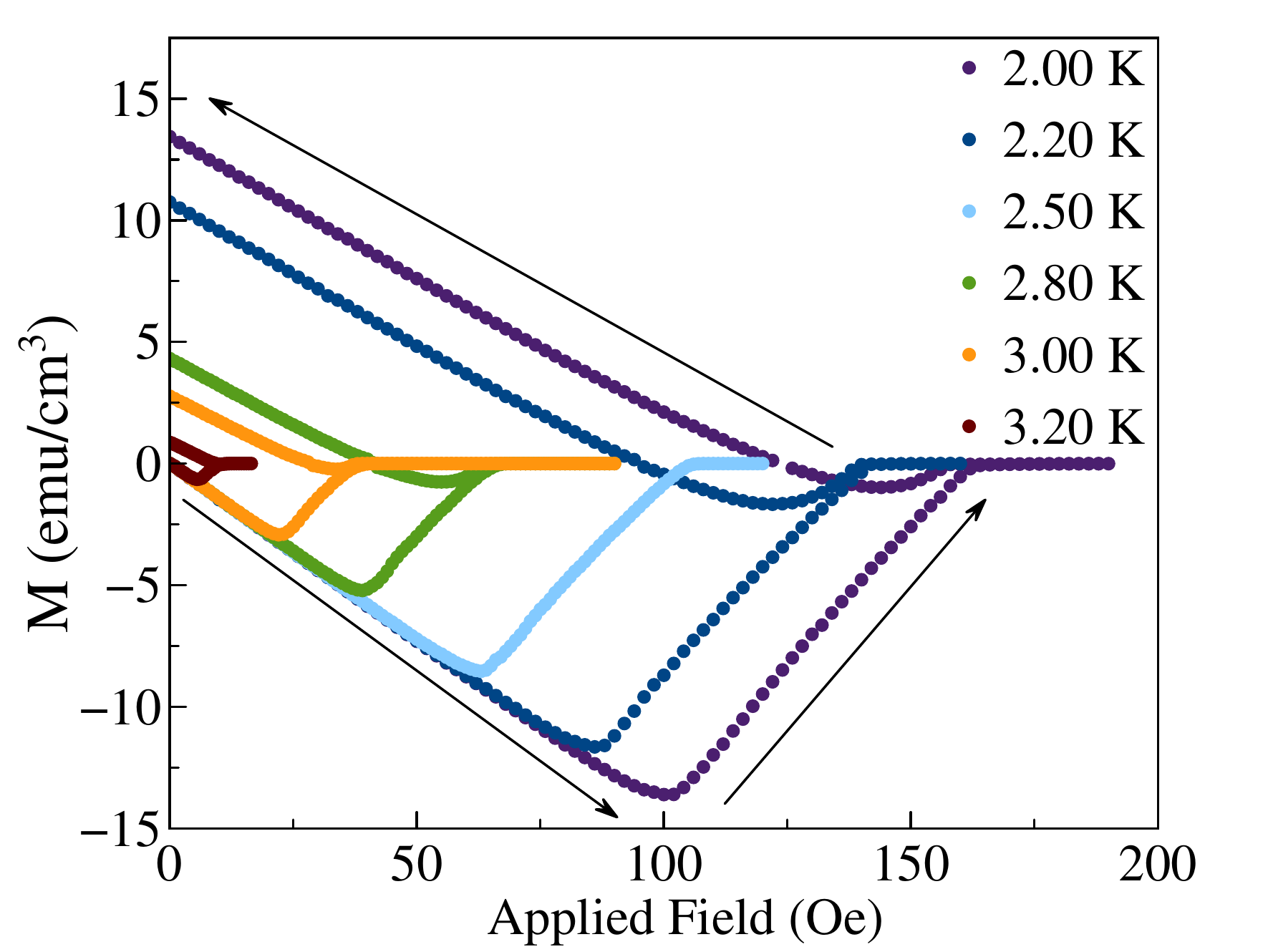}}
\subfigure[\label{fig:MvsHeq}]{
\includegraphics[width=0.37\textwidth]{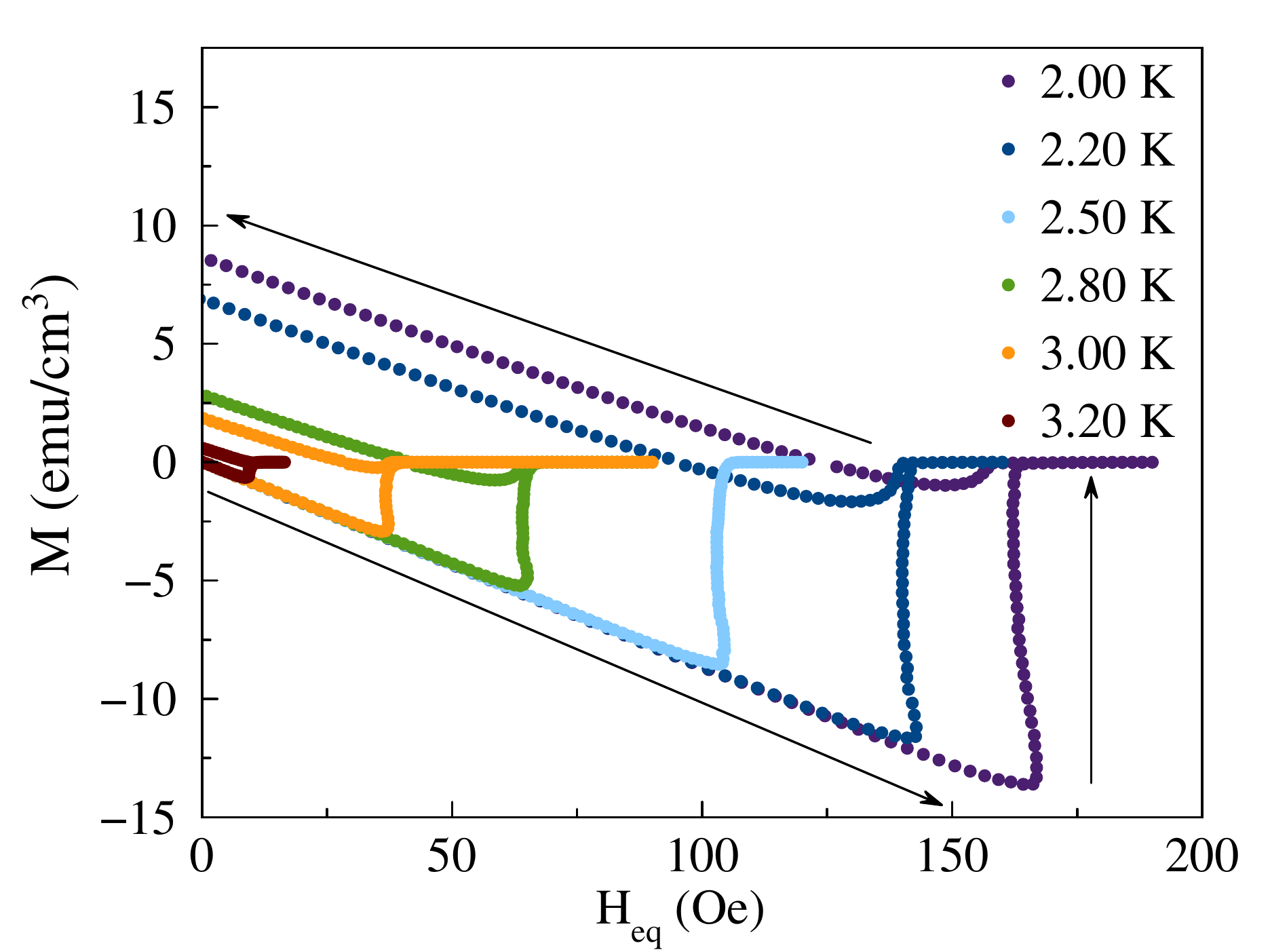}}
\caption{\label{fig:fig4} (a) Magnetisation as a function of applied field from 2.0K to 3.2K. Measurements were taken in increasing field followed by measurements in decreasing field once H$_{eq} >$H$_{cI}$ was reached. (b) Magnetisation as a function of the field at the equator H$_{eq} =$ H$_{a} - 4\pi N M$. This plot shows the expected discontinuous transition in magnetisation for a type-I superconductor}
\end{figure}
H$_{eq}$ is an important quantity because while in the intermediate state the internal field is equal to the surface field on the equator of the superconductor. As this is the region with the highest local field, when the sample enters the normal state the equator should be the first region to do so. By using the fact that H$_{eq}$ = H$_{c}$ while in the intermediate state, we should be able to reconstruct the discontinuous magnetisation behaviour expected at H$_{c}$ for a type-I superconductor. An ellipsoidal type-I superconductor will be in the Meissner state below $H_{a} \approx \left( 1 - N \right) H_{cI}$ above which it will enter the intermediate state\cite{Tinkham1975,deGennes1999}. Fig. \ref{fig:MvsHRaw} shows the magnetisation as a function of applied field for temperatures from 0.5K to 3.2K. Measurements were taken from 0 G to a maximum field, followed by measurements from the maximum field to 0 G. Fields are accurate to 0.1 G. A linear relationship is seen at low fields, with a slight departure from linear as the minima is approached, after which the magnetisation increases linearly until a transition is reached. This behaviour is well described by a type-I superconductor entering the intermediate state near the magnetisation minima followed by a transition to the normal state at high fields. The departure from linearity as the minima is approached is due to the generation of normal regions and the restructuring of normal-superconducting domains as the sample enters the intermediate state \cite{Landau1948,Fortini1972}. 

H$_{eq}$ has very different behaviours for type-I superconductors in the intermediate state and type-II superconductors in the vortex state. In the intermediate state, the superconducting volume fraction of the sample decreases approximately linearly for a wide range of applied fields while the microscopic magnetisation does not change. The linear decrease in the superconducting volume makes the overall magnitude of the magnetisation of the entire sample decrease linearly. H$_{eq}$ therefore remains constant, equal to internal field while in the intermediate state of a type-I superconductor.

A demagnetising factor of N$_{ellipsoid}$ = 0.4355 was calculated using Eq. 34 of Ref. \cite{Beleggia2006} assuming the sample was a perfect ellipsoid with major-axis a = b = 2.75 mm and minor-axis c = 1.90 mm. Fig. \ref{fig:MvsHeq} shows the magnetisation as a function of H$_{eq}$ demonstrates a discontinuous transition in magnetisation that is expected for a type-I superconductor \cite{deGennes1999,Tinkham1975}. The sample is not a perfect ellipsoid and a demagnetising factor of N$_{sample} =$ 0.3755 was found by optimizing the discontinuous transition of the magnetisation in Fig. \ref{fig:MvsHeq}, such that the transition occurred over the smallest H$_{eq}$ range. A high quality sample of a type-I superconducting material should exhibit a crisp transition at H$_{cI}$. The high quality nature of our sample \cite{Amon2018}, along with our $\mu$SR measurements showing type-I behaviour, justify this optimisation procedure for the demagnetisation factor. Fig. \ref{fig:MvsHeq} shows that in the intermediate state the magnetisation as a function of H$_{eq}$ is nearly vertical. The change in H$_{eq}$ as the magnitude of the magnetisation decreases is due to the thermodynamic critical field being modified from H$_{c}$ at low field to H$_{cI}$ at fields just below H$_{cI}.$

\begin{figure}[!h]
\centering
\includegraphics[width=0.37\textwidth]{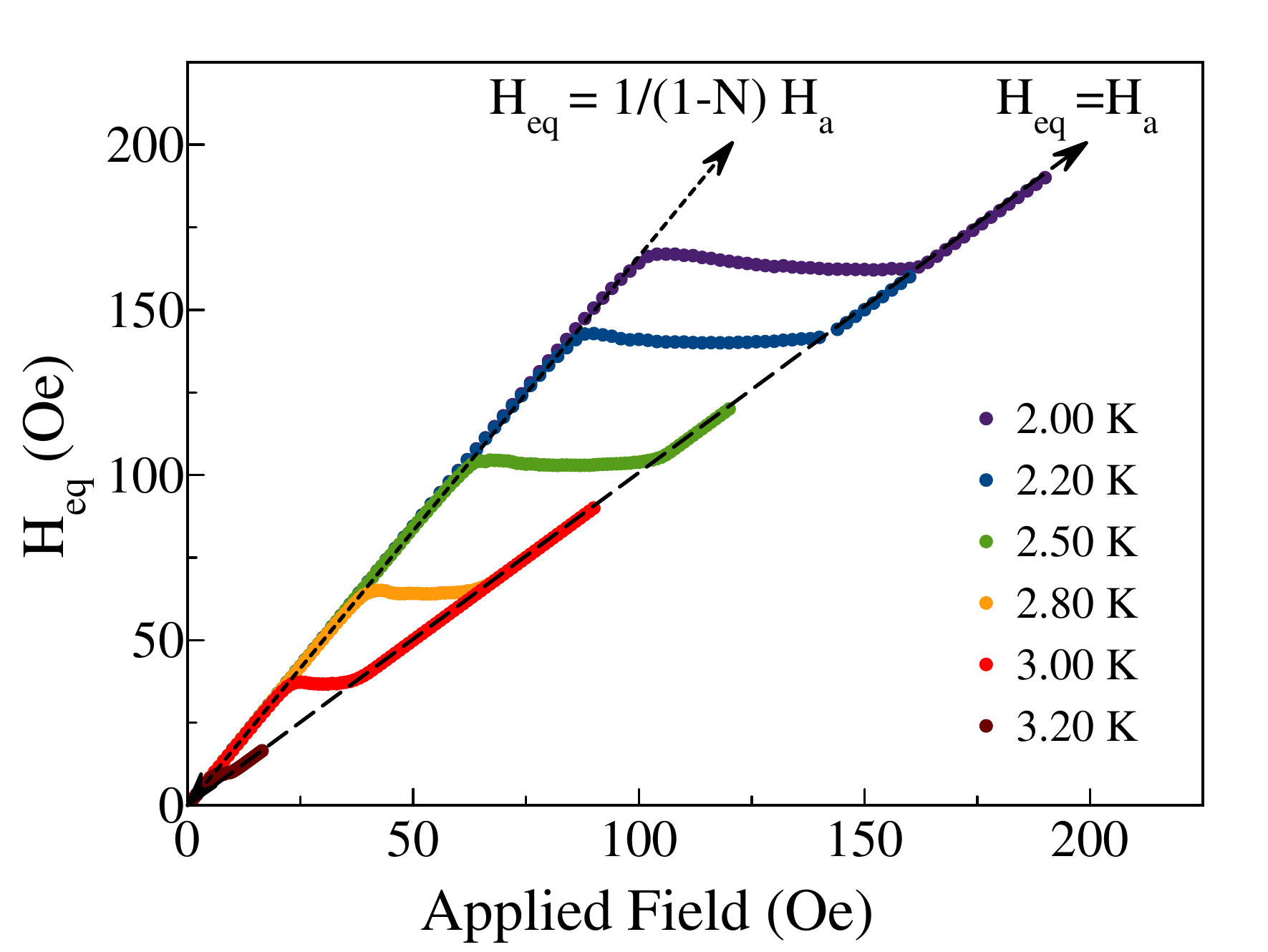}
\caption{\label{fig:HeqVsHa} Internal Field at the equator as a function of applied field. In the Meissner State, H$_{eq}$ = $\frac{1}{1 - N}H_{a}$ and the transition to the Intermediate State can be identified for applied fields where H$_{eq}$ becomes approximately constant with H$_{cI} \leq$ H$_{eq} \leq$ H$_{c}$. In the Normal State H$_{eq}$ = H$_{a}$.}
\end{figure}

Fig. \ref{fig:HeqVsHa} shows the field at the equator, H$_{eq}$, of the sample as a function of applied field. This plot matches the expected behaviour of H$_{eq}$ for a superconducting ellipsoid given by \cite{Tinkham1975, deGennes1999} where there is a linear increase in H$_{eq}$ while in the Meissner state with H$_{eq}$ = $\frac{H_{a}}{1 - N}$ as H$_{a}$ is increased to (1 - N) H$_{cI}$. In the intermediate state H$_{eq}$ stays relatively constant with some modification due to the thermodynamic critical field changing from H$_{c}$ at low fields to H$_{cI}$ near H$_{cI}$. Above H$_{cI}$, H$_{eq}$ increases linearly with H$_{eq}$ = H$_{a}$.

\begin{figure}[!h]
\centering
\subfigure[\label{fig:ChiVsT}]{
\includegraphics[width=0.37\textwidth]{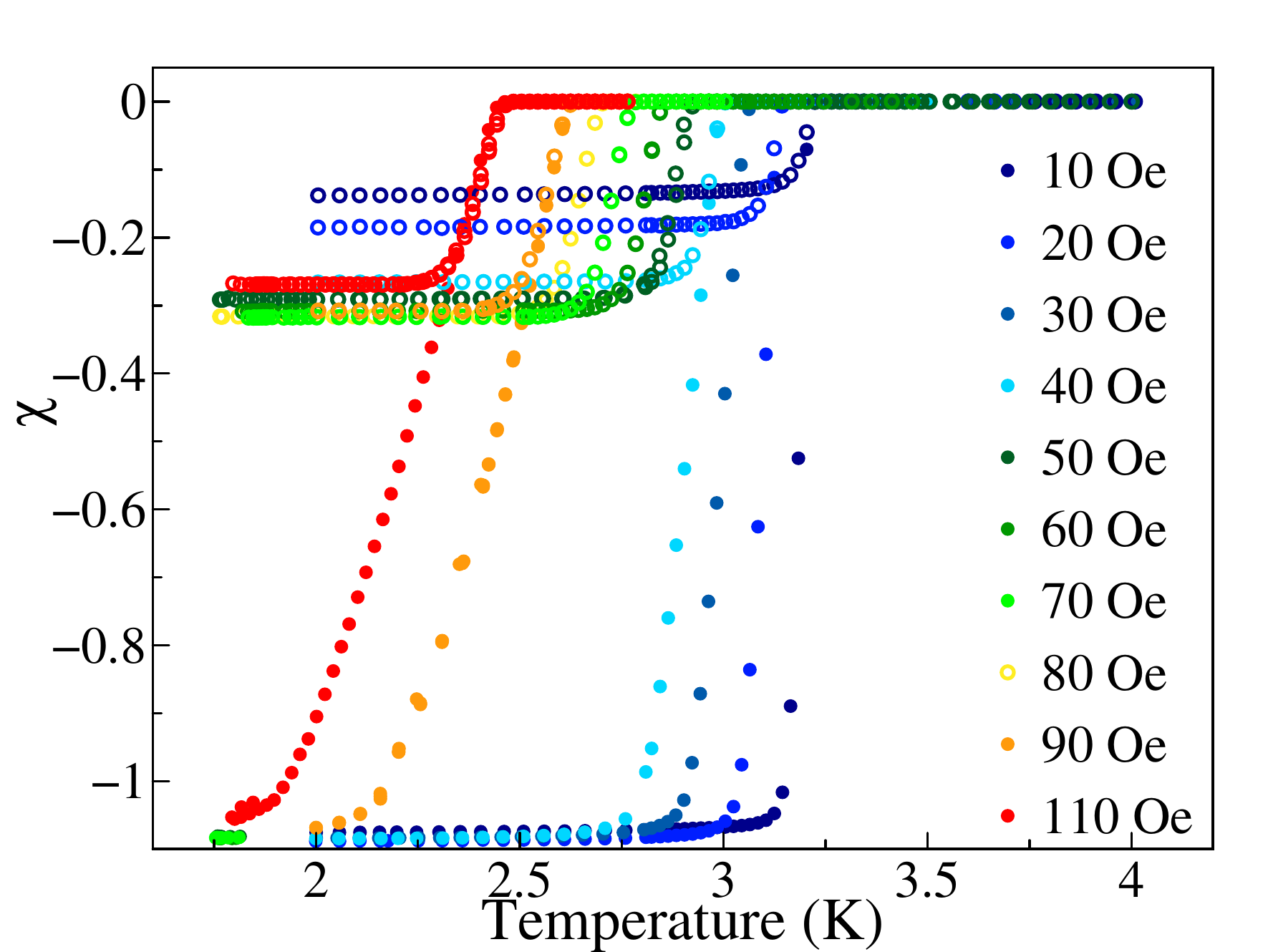}}
\subfigure[\label{fig:HeqVsT}]{
\includegraphics[width=0.37\textwidth]{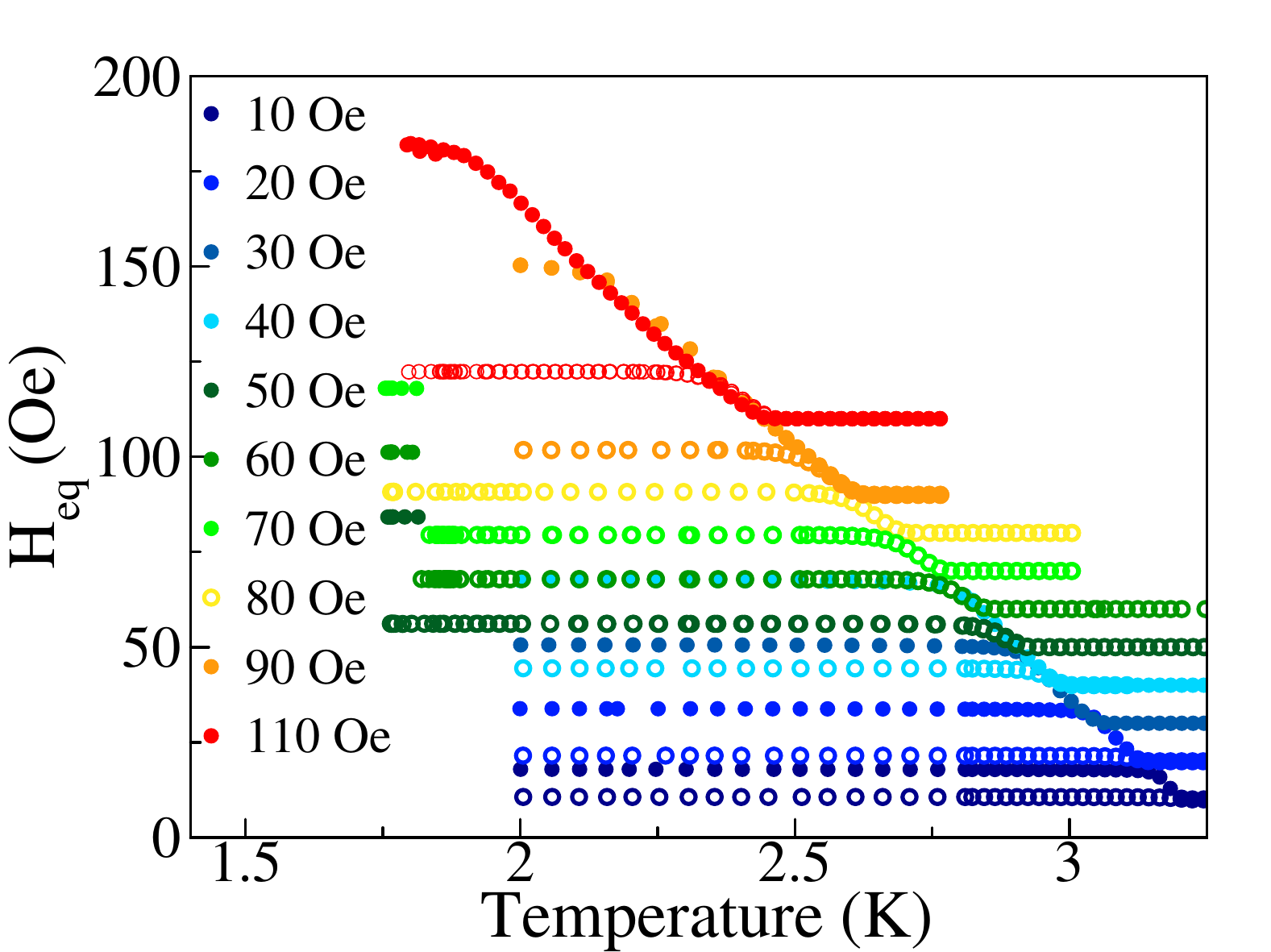}}
\subfigure[\label{fig:HcVsT_Chi}]{
\includegraphics[width=0.37\textwidth]{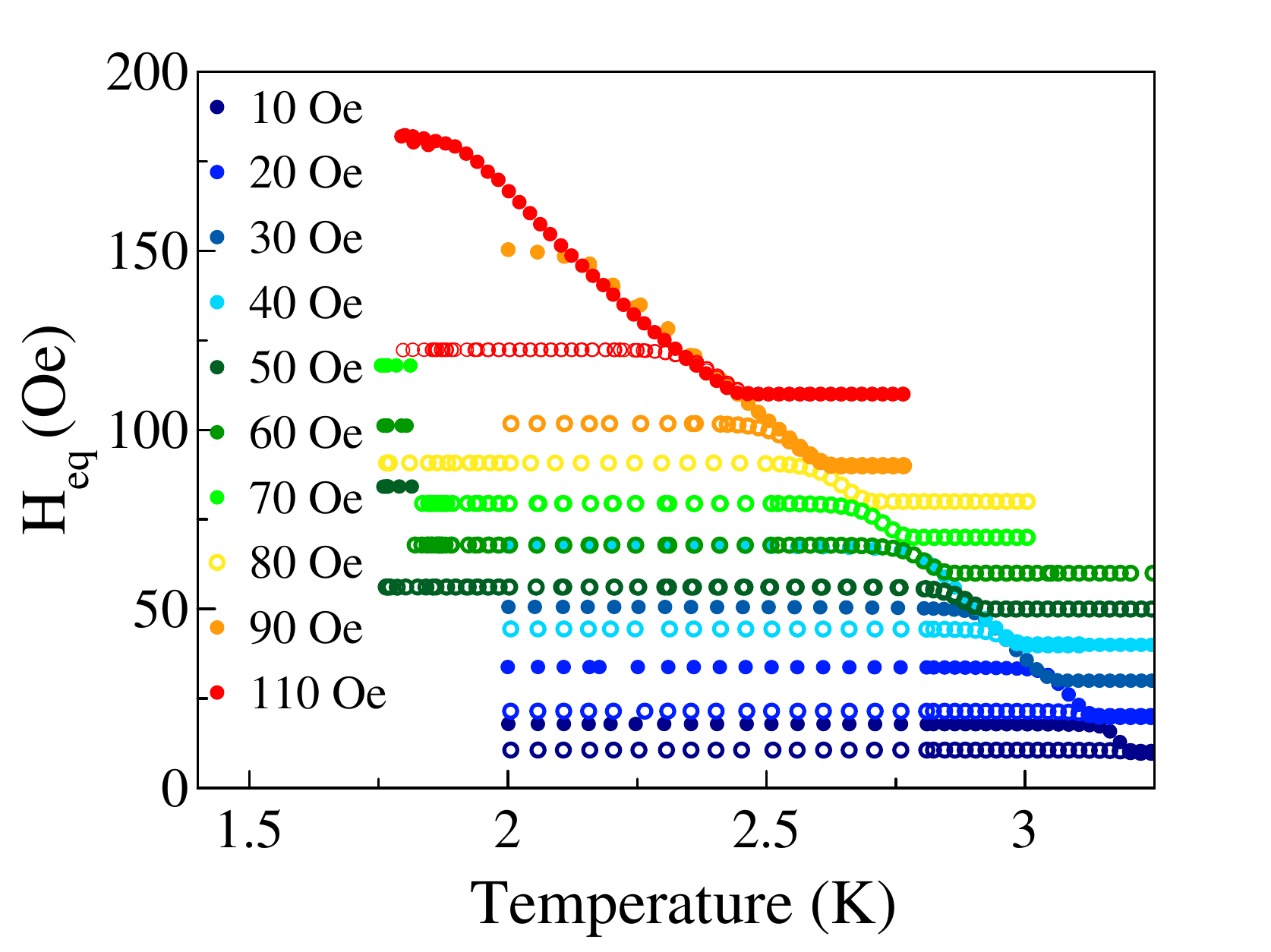}}
\caption{\label{fig:fig5} (a) Magnetic Susceptibility as a function of temperature for applied fields from 10-110 Oe. Zero field cooled (ZFC) (closed circles) and field cooled (FC) (open circle) data show small overlap regions. (b) Equatorial field as a function of temperature for applied fields from 10-110 Oe. ZFC (closed circles) and FC (open circles) show there are small overlap regions. (c) H$_{c}$(T) may be mapped out by using data points where the ZFC and FC regions overlap and is fit to Eq. \ref{Eq:GL} yielding H$_{c}$ = 258.6 $\pm$ 0.5 Oe and T$_{c}$ = 3.234 $\pm$ 0.003 K}
\end{figure}

The magnetic susceptibility in applied fields from 10-110 Oe was also measured as a function of temperature using the demagnetisation factor found from our previous measurements and is shown in Fig. \ref{fig:ChiVsT}. The zero field cooled (ZFC) measurements (closed circles) show a full magnetic flux expulsion with $\chi$ approaching (just below) -1 as T approaches zero, while the field cooled (FC) data (open circles) indicate there is some field being maintained in the sample. Small overlap regions in the intermediate state can be seen where the ZFC and FC results agree. H$_{eq}$ as a function of temperature is shown in Fig. \ref{fig:HeqVsT} again showing small regions of agreement between ZFC and FC. In these small regions of reversibility H$_{cI}$ can be mapped out as a function of temperature as shown in Fig. \ref{fig:HcVsT_Chi} which can be fit to the Ginzburg-Landau relation,
\begin{equation}\label{Eq:GL}
H_{c}\left(T\right) = H_{c}(0)\left(1-\left(\frac{T}{T_{c}}\right)^{2}\right),
\end{equation}
\cite{Tinkham1975,deGennes1999} yielding H$_{c}$(0) = 258.6 $\pm$ 0.5 Oe and T$_{c}$ = 3.234 $\pm$ 0.003 K.

\begin{figure}[!h]
\centering
\includegraphics[width=0.37\textwidth]{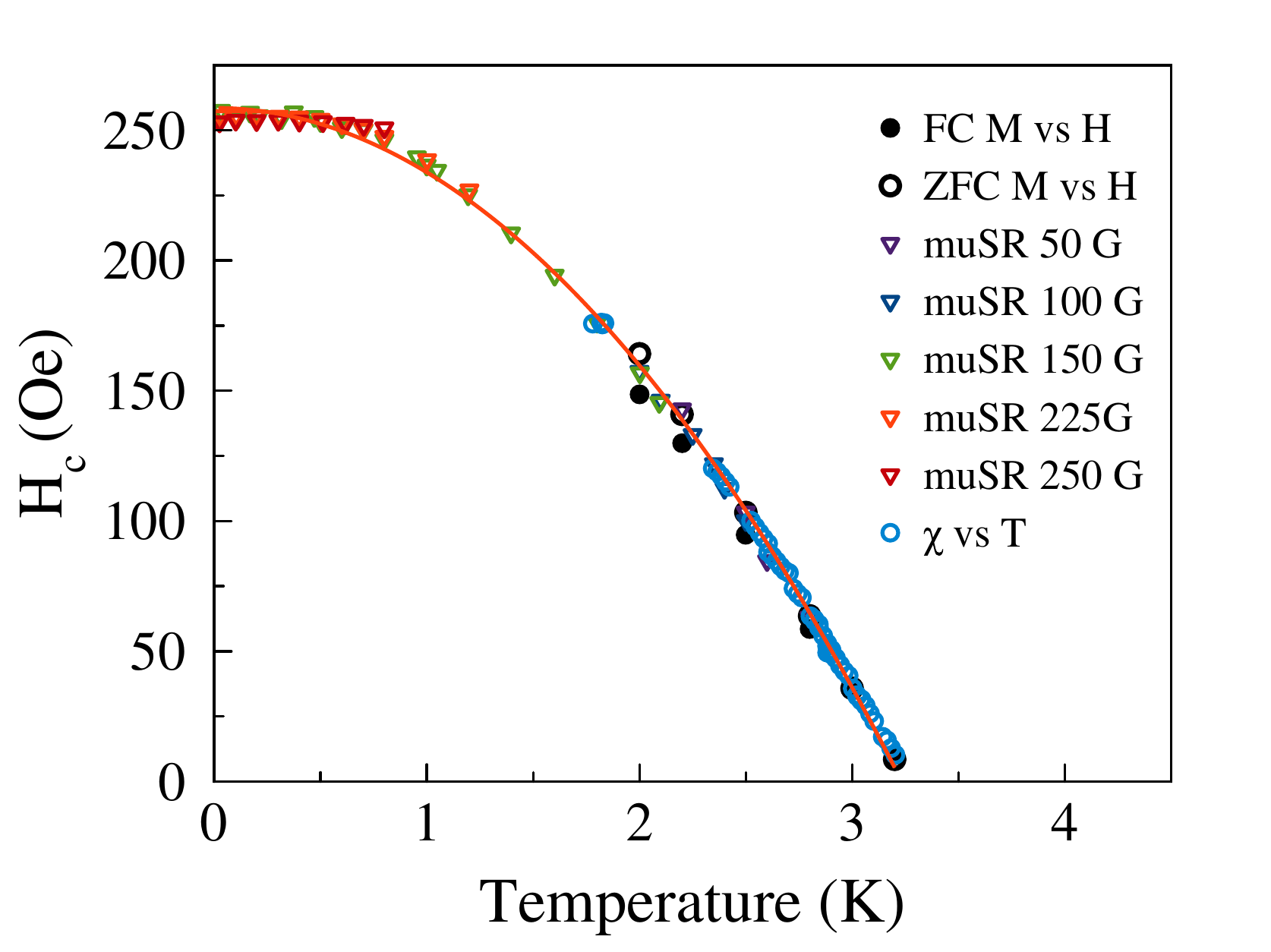}
\caption{\label{fig:PhaseDiagram} The Field versus Temperature phase diagram combining together our $\mu$SR (open triangles), magnetisation (black circles) and magnetic susceptibility (blue open circles) measurements. The diagram shows H$_{c}$(0)$\approx$ 256 Oe and T$_{c} \approx$  3.25 K}
\end{figure}

Combining the results of our $\mu$SR measurements (triangles), magnetisation versus applied field (black circles) and magnetic susceptibility versus temperature (blue open circles) yields the phase diagram shown in Fig. \ref{fig:PhaseDiagram}. Overlap regions between $\mu$SR and magnetisation measurements show good agreement.  A fit to Eq. \ref{Eq:GL} shows good agreement with the experimental data, with the low temperature $\mu$SR data having a slightly flatter than quadratic behaviour at low temperature.

\begin{figure}[!h]
\centering
\includegraphics[width=0.37\textwidth]{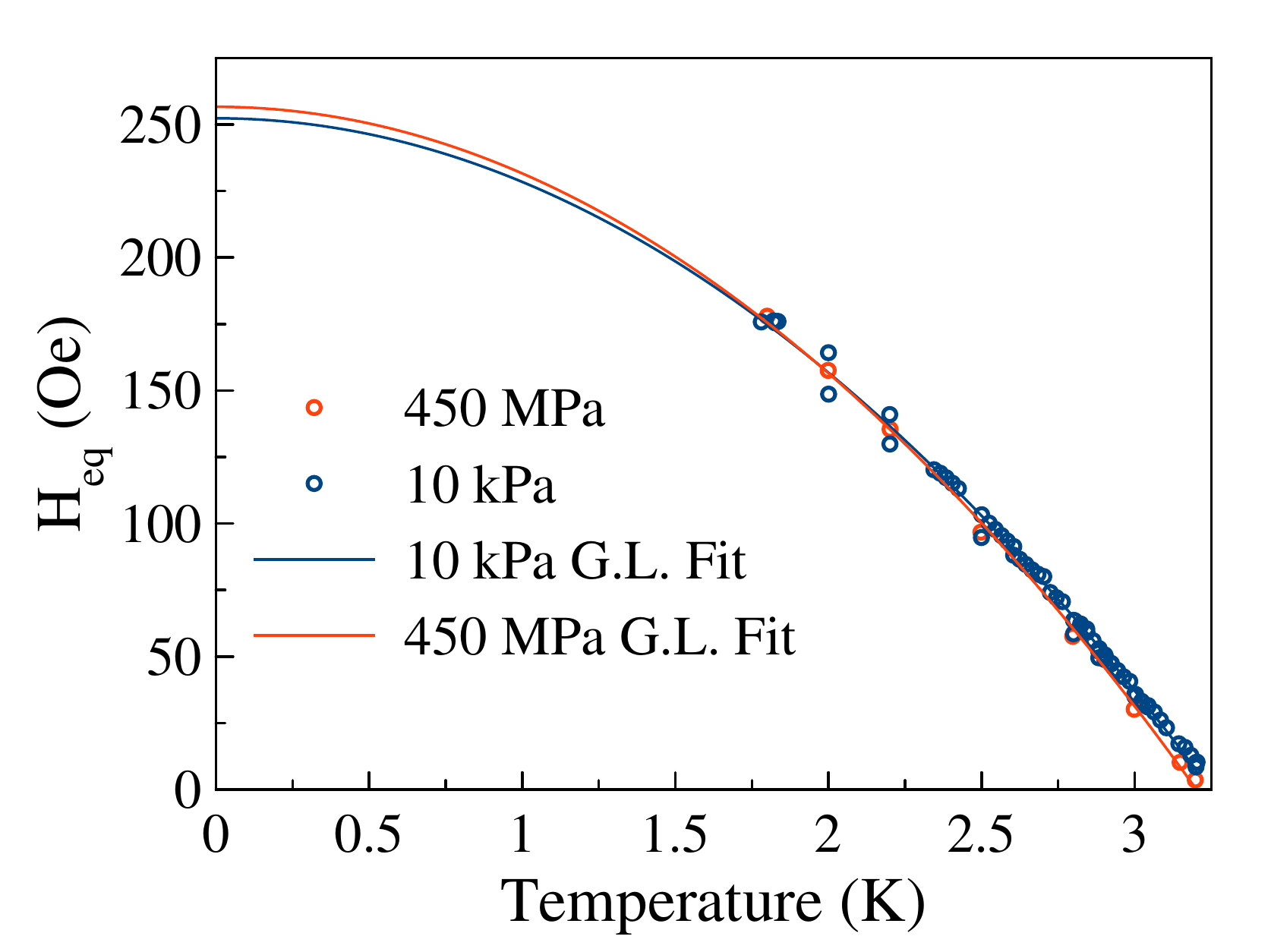}
\caption{\label{fig:PressurePhaseDiagram} The magnetic field versus temperature phase diagram from our magnetometry data at 10 kPa (blue circles) and 450 MPa (red circles). A Ginzburg-Landau fit (Eq. \ref{Eq:GL}) to the data between  yields H$_{c}$ = 258.6 $\pm$ 0.5 Oe (257 $\pm$ 3 Oe) and T$_{c}$ = 3.234 $\pm$ 0.003 K (3.20 $\pm$ 0.01 K) for the 10 kPa (450 MPa) data.}
\end{figure}

The effect of pressure on the system was explored using a pressure cell inserted into the MPMS. Fig. \ref{fig:PressurePhaseDiagram} shows the results for ambient pressure ($\approx$10kPa, blue circles) and 450 MPa (red circles) as well as fits to the ambient pressure (blue line) and 450 MPa (red line) data. Using Eq. \ref{Eq:GL}, at 450 MPa, H$_{c}$ = 257 $\pm$ 3 Oe and T$_{c}$ = 3.20 $\pm$ 0.01 K, a change in T$_{c}$ of 34 $\pm$ 11 mK compared to the 10 kPa pressure data with H$_{c}$ consistent with the 10 kPa data. The change in T$_{c}$ is comparable to the change observed in the elemental type-I superconductor Tin (IV), Indium and Tantalum which have a decrease in T$_{c}$ of about 20 mK under the same conditions \cite{Jennings1959}. This suggests that BeAu is far from a quantum critical point accessible by application of pressure.

\section{\label{sec:level1}Conclusion}

$\mu$SR and demagnetisation corrected magnetisation measurements were carried out on discs and an ellipsoid of polycrystalline BeAu. Our results show that BeAu is a type-I superconductor with H$_{c}\approx$ 256 Oe and T$_{c} \approx$3.25 K. The $\mu$SR and magnetisation results show consistent values of H$_{c}$(T) in the regions where they overlap. A Ginzburg-Landau fit (Eq. \ref{Eq:GL}) gives H$_{c}$ = 258.6 $\pm$ 0.5 Oe and T$_{c}$ = 3.234 $\pm$ 0.003 K. Magnetisation measurements on an ellipsoid were taken in a pressure chamber at 450 MPa. Fitting the data to a Ginzburg-Landau (Eq. \ref{Eq:GL}) relation yields H$_{c}$ = 257 $\pm$ 3 Oe and T$_{c}$ = 3.20 $\pm$ 0.01 K for the 450 MPa data and H$_{c}$ 258.6 $\pm$ 0.5 Oe and T$_{c}$ 3.234 $\pm$ 0.003 K for the 10 kPa pressure data. The reduction in T$_{c}$ under 450 MPa of pressure in BeAu is comparable to the decrease in T$_{c}$ observed in the elemental type-I superconductors Tin (IV), Indium and Tantalum under the same conditions \cite{Jennings1959}. This suggests that BeAu is far from a quantum critical point accessible by the application of pressure.

During the preparation of this manuscript we became aware of a preprint \cite{Singh2019} that reports similar results on BeAu.

\section{\label{sec:level1}Acknowledgements}
We thank G. D. Morris, B. S. Hitti, and D. J. Arseneau for their assistance with the $\mu$SR measurements. We thank Paul Dube for their assistance with the magnetometry measurements at McMaster. Work at McMaster University was supported by the Natural Sciences and Engineering Research Council of Canada and the Canadian Foundation for Innovation. ES greatly appreciates the support provided by the Fulbright Canada Research Chair Award.

\newpage
\bibliography{BeAu_Manuscript_v1}
\bibliographystyle{apsrev4-1}
\end{document}